\documentclass[twocolumn]{aastex6}
\usepackage{natbib}
\usepackage{color}
\usepackage{hyperref} 
\usepackage{booktabs}
\usepackage{mhchem}

\def    \apjl  		{\rm {ApJL}}
\def    \apj  		{\rm {ApJ}}
\def    \mnras  	{\rm {MNRAS}}
\def    \araa  		{\rm {ARA\& A}}
\def    \apjl  		{\rm {ApJL}}

\def	\cm		{\,{\rm {cm}}}
\def	\K		{\,{\rm K}}
\def	\g		{\,{\rm {g}}}

\def \bea {\begin{eqnarray}}
\def \ena {\end{eqnarray}}

\def	\cm	{\,{\rm cm}}
\def	\d	{{\rm d}}

\def	\erg	{\,{\rm erg}}

\def	\g	{\,{\rm g}}
\def	\gas	{\,{\rm gas}}

\def	\H	{{\rm H}}

\def    \COMs {{\rm COMs}}

\def	\disr 	{\rm disr}
\def	\max 	{\rm max}

\usepackage{amsmath}

\begin{document}

\shorttitle{Rotational desorption in Orion BN/KL}
\shortauthors{Tram et al.}
\title{Observational evidence for rotational desorption of Complex Molecules by radiative torques from Orion BN/KL}

\author{Le Ngoc Tram\altaffilmark{1}, Hyeseung Lee\altaffilmark{2}, Thiem Hoang\altaffilmark{2,}\altaffilmark{3}, Joseph M. Michail\altaffilmark{4}, David T. Chuss\altaffilmark{5}, Sarah Nickerson\altaffilmark{6}, Naseem Rangwala\altaffilmark{6}, William T. Reach\altaffilmark{1}}

\affil{$^1$ Stratospheric Observatory for Infrared Astronomy, Universities Space Research Association, NASA Ames Research Center, MS 232-11, Moffett Field, 94035 CA, USA; \href{mailto:ngoctram.le@nasa.gov}{ngoctram.le@nasa.gov}}
\affil{$^2$ Korea Astronomy and Space Science Institute, Daejeon 34055, South Korea}
\affil{$^3$ Korea University of Science and Technology, 217 Gajeong-ro, Yuseong-gu, Daejeon, 34113, South Korea}
\affil{$^4$ Center for Interdisciplinary Exploration and Research in Astrophysics (CIERA) and Department of Physics and Astronomy Northwestern University, 1800 Sherman Ave Evanston, IL 60201, USA}
\affil{$^5$ Department of Physics, Villanova University, 800 E. Lancaster Ave., Villanova, PA 19085, USA}
\affil{$^6$ NASA Ames Research Center, Moffett Field, CA 94035, USA}

\begin{abstract}
Complex Organic Molecules (COMs) are believed to form in the ice mantle of dust grains and are released to the gas by thermal sublimation when grain mantles are heated to temperatures of $T_{\d}\gtrsim 100\K$. However, some COMs are detected in regions with temperatures below 100 K. Recently, a new mechanism of rotational desorption due to centrifugal stress induced by radiative torques (RATs) is proposed by \cite{2020ApJ...891...38H} that can desorb COMs at low temperatures. In this paper, we report observational evidence for rotational desorption of COMs toward the nearest massive star-forming region Orion BN/KL. We compare the abundance of three representative COMs which have very high binding energy computed by the rotational desorption mechanism with observations by ALMA, and demonstrate that the rotational desorption mechanism can explain the existence of such COMs. We also analyze the polarization data from SOFIA/HAWC+ and JCMT/SCUBA-2 and find that the polarization degree at far-infrared/submm decreases with increasing the grain temperature for $T_{\d}\gtrsim 71\K$. This is consistent with the theoretical prediction using the Radiative Torque (RAT) alignment theory and Radiative Torque Disruption (RATD) mechanism. Such an anti-correlation between dust polarization and dust temperature supports the rotational disruption as well as rotational desorption mechanism of COMs induced by RATs.
\end{abstract}

\section{Introduction} \label{sec:intro}
Complex organic molecules (or COMs) are the building blocks of life. 
COMs are believed to first form on the icy mantle of dust grains (e.g., \citealt{2008ApJ...682..283G}; \citealt{2016ApJ...830L...6J}). Subsequently, COMs are released to the gas phase during the warm and hot phase, where icy grain mantles can be heated to high temperatures $T_{\d}\sim 100 - 300\K$ (e.g., \citealt{1987ApJ...315..621B}; \citealt{1988MNRAS.231..409B}; \citealt{2007A&A...465..913B}).
COMs are increasingly observed in the environs of young stellar objects (YSOs),
including hot cores/corinos around high-mass/low-mass
protostars and protoplanetary disks (see \citealt{2009ARA&A..47..427H} and \citealt{2014FaDi..168....9V} for recent reviews). 
In particular, many surveys that search for COMs toward 
the nearest massive star-forming regions, the Orion \textit{Becklin-Neugebauer/Kleinmann-Low} object (BN/KL), have been conducted with
single-dish telescopes from the ground, 
e.g., IRAM-30m (\citealt{2010A&A...517A..96T, 2011A&A...528A..26T};
\citealt{2016A&A...587L...4C}), in space, 
e.g., Herschel (\citealt{2010A&A...521L..20B}; \citealt{2014ApJ...787..112C}), 
and with interferometers, e.g., IRAM Plateau de Bure (\citealt{2011A&A...532A..32F}; \citealt{2012A&A...543A.152P}; \citealt{2013A&A...550A..46B}), and ALMA (\citealt{2017ApJ...837...49P, 2019ApJ...871..251P}; \citealt{2015A&A...582L...1T,2018A&A...620L...6T}; \citealt{2017A&A...604A..32P, 2019A&A...624L...5P}; \citealt{2017A&A...604L...2F}).
The single-dish telescopes are unable to spatially/spectrally separate the contribution of the different sources in Orion BN/KL because of their low resolution (i.e., lines are overlapped). All constraints from those data, therefore, are manifested as upper limits.  
Conversely, the extremely high spatial resolution of interferometers naturally can separate the components, which reveals the complexity of the chemical
and physical structure in Orion BN/KL. Observations show the existence of large COMs with very strong binding energies, such as acetic acid (CH$_{3}$COOH), ethyl formate (C$_{2}$H$_{5}$OCHO), propyl cyanide (C$_{3}$H$_{7}$CN), methyl acetate (CH$_{3}$COOCH$_{3}$), methoxymethanol (CH$_{3}$OCH$_{2}$OH), ethylene glycol (OHCH$_{2}$CH$_{2}$OH), etc. The sublimation temperature of these large COMs is higher than that of water, i.e., $T_{\rm sub}\gtrsim 152$ K (see Section \ref{sec:COM_obs}). However, the dust temperature estimated in Orion BN/KL is far from $T_{\rm sub}$. This casts doubt on the popular sublimation mechanism and suggests non-thermal mechanisms that can desorb COMs within these environments.

\cite{2019NatAs...3..766H} discovered that large dust grains exposed in the intense radiation (e.g., in massive star-forming regions) are spun-up to extremely fast rotation due to Radiative Torques (RATs) such that the centrifugal stress can exceed the tensile strength of the grain material, breaking the original grain into many fragments. This is referred to as the RAdiative Torques Disruption (RATD) mechanism. Then, \cite{2020ApJ...891...38H} demonstrated that RATD is efficient in disrupting the ice mantle from the grain core into small ice fragments when the radiation field is sufficiently intense. Subsequently, the evaporation of molecules from tiny fragments can occur through transient heating, or enhanced thermal sublimation due to the increase of the grain temperature with decreasing the grain size. 
As a result, COMs could be released to the gas before the grains can be heated to their sublimation temperatures. This mechanism is so-called the rotational desorption, and numerical calculations show this mechanism can be efficient at temperatures much below the sublimation temperature required to desorb COMs (see Figure 7 in \citealt{2020ApJ...891...38H}). The grain suprathermal rotation by RATs is also found to accelerate thermal sublimation of molecules from the grain surface \citep{2019ApJ...885..125H}.                

\textit{The first goal of this work is to study if the rotational desorption can explain the detection of numerous COMs quantitatively in the massive-star formation region in Orion BN/KL by ALMA observations}. 

In light of rotational desorption, \cite{2020ApJ...891...38H} predicted a correlation of the abundance of COMs with the depletion of large grains in dense regions with intense radiation such as the Orion BN/KL. Such a depletion arises both from the removal of the ice mantles from the grain core or the disruption of the entire grains into fragments. Observationally, the decrease of large grains is expected to decrease the polarization degree of thermal dust emission at far-infrared/submm because the polarization at these wavelengths is most sensitive to large grains (see \citealt{2020Galax...8...52H} for a review). It is worth to mention that dust polarization arises from the alignment of dust grains with the magnetic field. After seven decades of studying of grain alignment, the RAT alignment has become the most popular theory to explain grain alignment (see \citealt{2008MNRAS.388..117H, 2016ApJ...831..159H}; \citealt{2015ARA&A..53..501A}). One of the key predictions of the RAT alignment theory is that the degree of grain alignment depends on the local conditions, including the radiation field and gas properties. The RAT theory implies an increase in the polarization fraction with increasing the radiation strength, which is numerically demonstrated through modeling in \citealt{2020ApJ...896...44L}. Moreover, when the radiation is sufficiently high, such that RATD happens, the dust polarization fraction at long wavelengths decreases due to the conversion of large grains into small ones (see \citealt{2020ApJ...896...44L} and \citealt{2020arXiv200710621T} for the details). {\it The second goal of this paper is to test the correlation of COMs and dust polarization with dust temperature from the Orion BN/KL}.

We use dust polarization data observed toward the Orion BN/KL by the High-resolution Airborne Wideband Camera Plus (HAWC+) polarimetric instrument (\citealt{2018JAI.....740008H}) onboard Stratosphere Observatory for Infrared Astronomy (SOFIA) and the Submillimetre Common-User Bolometer Array 2 (SCUBA-2) accommodated on the James Clerk Maxwell Telescope (JCMT) (see \citealt{2016MNRAS.461.4022M}; \citealt{2017ApJ...846..122P}; \citealt{2017ApJ...842...66W}; \citealt{2019ApJ...872..187C}). We infer dust temperature using the high-resolution dust map of this area from the Faint Object infraRed CAmera for the SOFIA Telescope (FORCAST) (\citealt{2010SPIE.7735E..1UA}).

The remainder of this paper is structured as follows. We present observational information toward Orion BN/KL in Section \ref{sec:obs_info}. We briefly describe the rotational desorption mechanism and the application for the case of Orion BN/KL in Section \ref{sec:rot_desorp}. In Section \ref{sec:polarization_COMs}, we present our analysis of dust polarization observed by HAWC+ and SCUBA-2 and present a correlation between COMs and dust polarization, as well as our numerical modeling reproducing the observed anti-correlation between the dust polarization and temperature. We summarize our main findings in Section \ref{sec:summary}. 

\section{Observational information toward Orion BN/KL} \label{sec:obs_info}
\subsection{Detection of numerous large COMs} \label{sec:COM_obs}
As mentioned above, COMs are frequently observed in the nearest high-mass star-forming region Orion BN/KL. Interestingly, observations report the detection of many large COMs such as acetic acid (CH$_{3}$COOH), propyl cyanide (C$_{3}$H$_{7}$CN), ethyl formate (C$_{2}$H$_{5}$OCHO), methyl acetate (CH$_{3}$COOCH$_{3}$), methoxymethanol (CH$_{3}$OCH$_{2}$OH), ethylene glycol (OHCH$_{2}$CH$_{2}$OH), etc. (see e.g., \citealt{2017A&A...604A..32P, 2019A&A...624L...5P}; \citealt{2018A&A...620L...6T} and \citealt{2019ApJ...871..251P} for the recent reports). These large COMs have very high binding energies, i.e., E$_{\rm b}$(CH$_{3}$COOH)$\simeq$6300$\,$K, E$_{\rm b}$(C$_{3}$H$_{7}$CN)$\simeq$7240$\,$K, E$_{\rm b}$(C$_{2}$H$_{5}$OCHO)$\simeq$6250$\,$K, E$_{\rm b}$(CH$_{3}$OCH$_{2}$OH)=7580$\,$K, and E$_{\rm b}$(OHCH$_{2}$CH$_{2}$OH)=10200$\,$K (Table 4 in \citealt{2013ApJ...765...60G}). 
Comparing their binding energies to H$_{2}$O ($E_{\rm b}\simeq$5700$\,$K), one can infer that their sublimation temperatures should be higher than that of H$_{2}$O (T$_{\rm sub}(\rm H_{2}O)\simeq$152$\,$K, \citealt{1993prpl.conf.1177M}; \citealt{2015MNRAS.447.1444B}).

\subsection{Dust grain sizes}
Early polarization observations toward Orion (\citealt{1981ApJ...248..963B}; \citealt{1981ApJ...244..483M}; \citealt{1981MNRAS.194..485M}; \citealt{1999BASI...27..141V}) reported that the value of both the total-to-selective extinction ratio ($R_{V}$) and the wavelength at the maximum polarization ($\lambda_{\max}$) is relatively high in comparison to those in the interstellar medium (ISM) (i.e., $R_{V}\simeq 3.1$ and $\lambda_{\max}\simeq 0.55\,\mu$m). This indicates the existence of large grains of size of $a>0.2\,\mu$m. Furthermore, using a simple core-ice mantle model, \cite{1981MNRAS.194..485M} estimated the mean grain size about $0.26\,\mu$m - $0.29\,\mu$m for a core radius of $\lesssim 0.05\,\mu$m. 

\begin{figure}
    \centering
    \includegraphics[width=0.45\textwidth]{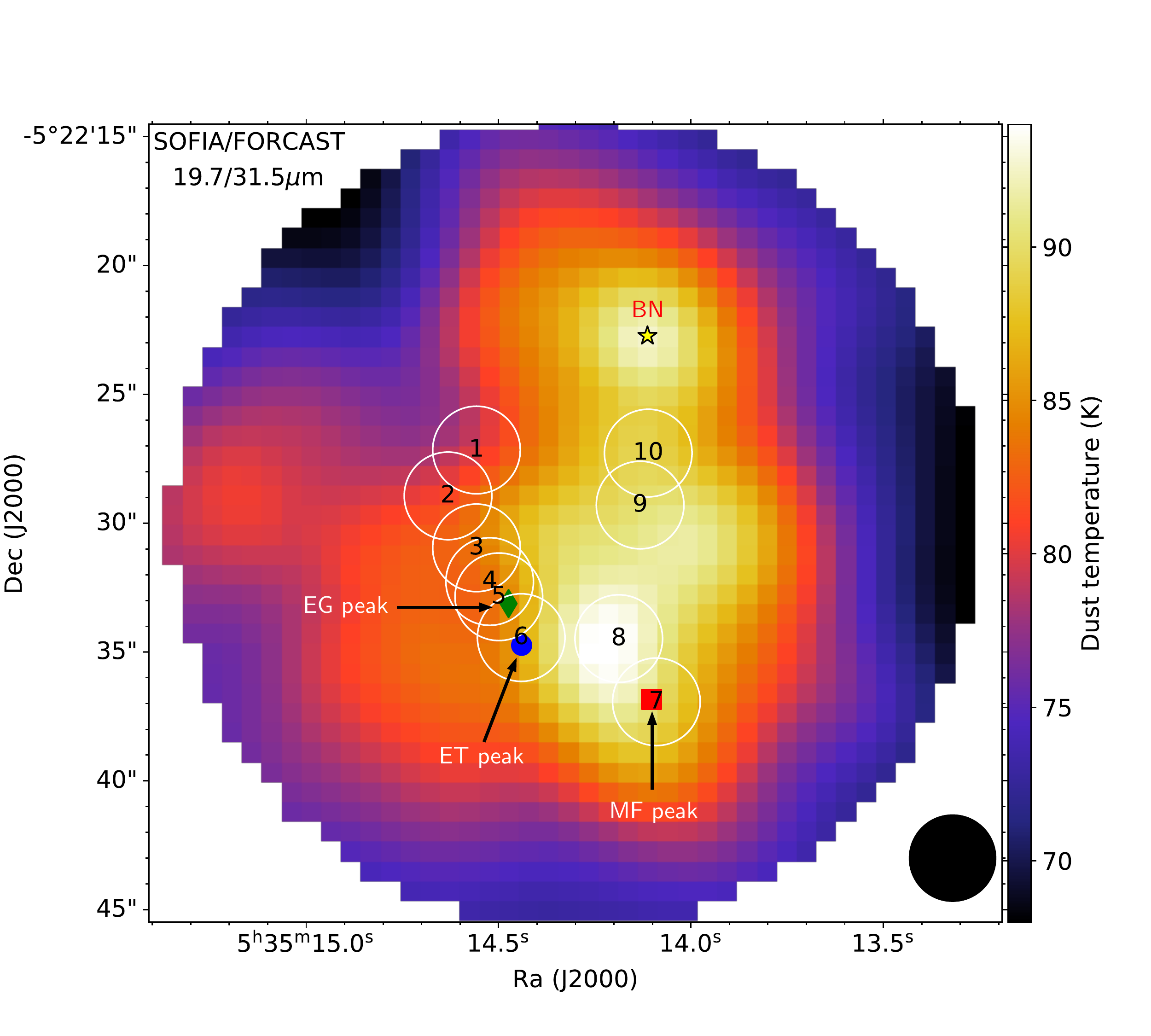}
        \caption{Map of dust temperature from SOFIA/FORCAST observations (\citealt{2012ApJ...749L..23D}). The beam size of SOFIA/FORCAST is $\simeq 3.5''$. 
        The numbers are the locations where ALMA has detected COMs (\citealt{2017A&A...604A..32P, 2019A&A...624L...5P}). The methyl formate (MF) peak, the ethylene glycol (EG) peak and the ethanol (ET) peak are denoted by the red square, the green diamond and the blue circle, respectively.}
        \label{fig:Td_map}
    \includegraphics[width=0.45\textwidth]{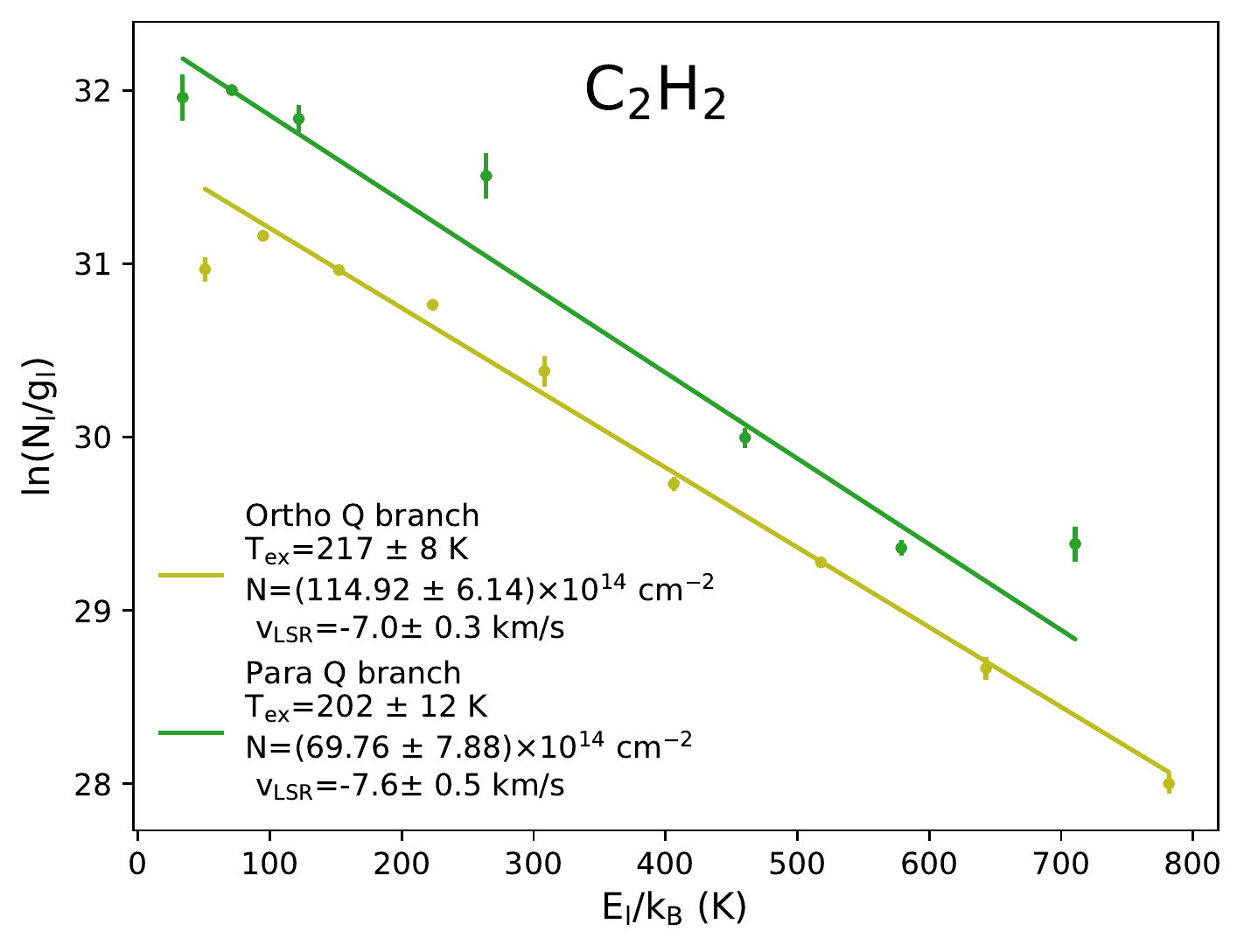}
    \caption{Rotational excitation diagram of C$_{2}$H$_{2}$ observed by SOFIA/EXES (\citealt{2018ApJ...856....9R}; \citealt{2020arXiv200812787N}). The diagram shows that the temperature of C$_{2}$H$_{2}$ is about $200\K$.}
    \label{fig:Tex}
\end{figure}
\subsection{Dust temperature}
\cite{2012ApJ...749L..23D} constructed a map of dust temperature using the early science data from FORCAST observations (\citealt{2010SPIE.7735E..1UA}) with a spatial resolution of $3.5''$ (\citealt{2012ApJ...749L..18H}). The construction is based on the 19.7/31.5 $\mu$m and 31.5/37.1 $\mu$m by assuming the optical depth relation $\tau_{19.7\,\mu \rm m}/\tau_{31.5 \,\mu \rm m}=3.7$ and $\tau_{31.5\,\mu \rm m}/\tau_{37.1\,\mu \rm m}=1.7$. This $3.5''$ resolution dust temperature map is shown in Figure \ref{fig:Td_map}. This map shows that the dust temperature in Orion BN/KL is lower than 100 K (i.e., the sublimation threshold). 

\cite{2019ApJ...872..187C} independently constructed a 18.7'' dust temperature map from combined data of SOFIA/HAWC+, Herschel/PACS, Herschel/SPIRE, JCMT/SCUBA-2, GBT/MUSTANG, GBT, and VLA observed toward Orion molecular cloud (OMC-1). This lower spatial resolution map shows that the dust temperature in Orion BN/KL is also lower than 100K. Note that the discrepancy between two these maps is not higher than 10 K.   

\subsection{Gas temperature and column density} \label{sec:Tex}
We use the value of the gas temperature derived from the rotational C$_{2}$H$_{2}$ ($J=9-8$ to $J=18-17$) transitions (\citealt{2018ApJ...856....9R}; Nickerson et al. in prep.). These lines are observed by the \textit{The Echelon-Cross-Echelle Spectrograph} (EXES; \citealt{2010SPIE.7735E..6QR}) onboard SOFIA with the resolution of $\sim 5\,\rm km\,s^{-1}$. Like molecular hydrogen, C$_{2}$H$_{2}$ has no permanent dipole moment; thus its transitions are optically thin, such that the excitation temperature could be built directly from observations (see Figure \ref{fig:Tex}). The slope of this diagram is proportional to the rotational excitation temperature $T_{\rm ex} \simeq 200\,\K$. This constraint agrees well with that obtained from the millimeter \ce{^12}CO(2-1) line by CARMA-NRO observations (\citealt{2018ApJS..236...25K}).
Furthermore, the gas density of Orion BN/KL is sufficiently high (see Section \ref{sec:numerical_results}) such that local thermodynamic equilibrium is applicable. Therefore, the gas and excitation temperatures are equivalent. In what follows, the gas temperature is adopted as $200\K$.

The Orion BN/KL is a dense medium, in which the gas column density is observationally constrained as order of 10$^{23}\cm^{-2}$ to 10$^{24}\cm^{-2}$. The constraint were made by the gas tracer ratio (e.g., \citealt{2014ApJ...795...13B}), line fitting (e.g., \citealt{2014ApJ...787..112C}) or dust continuum spectral energy distribution fitting (e.g., \citealt{2015ApJ...801...82H}).

\section{Modeling Rotational desorption in Orion BN/KL} \label{sec:rot_desorp} 
As for a demonstration, we apply the rotational desorption mechanism to explain the detection of three COMs: ethylene glycol (OHCH$_{2}$CH$_{2}$OH), methoxymethanol (CH$_{3}$OCH$_{2}$OH) and ethyl formate (C$_{2}$H$_{5}$OCHO) in this section. These molecules have very large binding energy and are observed toward different locations: at the methyl formate (MF) peak, the ethylene glycol (EG) peak and the ethanol (ET) peak in Orion BN/KL (see \citealt{2018A&A...620L...6T}). 

\subsection{Rotational desorption mechanism} 
The detailed mechanism of the rotational desorption mechanism is described in \cite{2020ApJ...891...38H}. In this section, we review the principles of this mechanism.

\begin{figure}[!ht]
    \centering
    \includegraphics[width=0.45\textwidth]{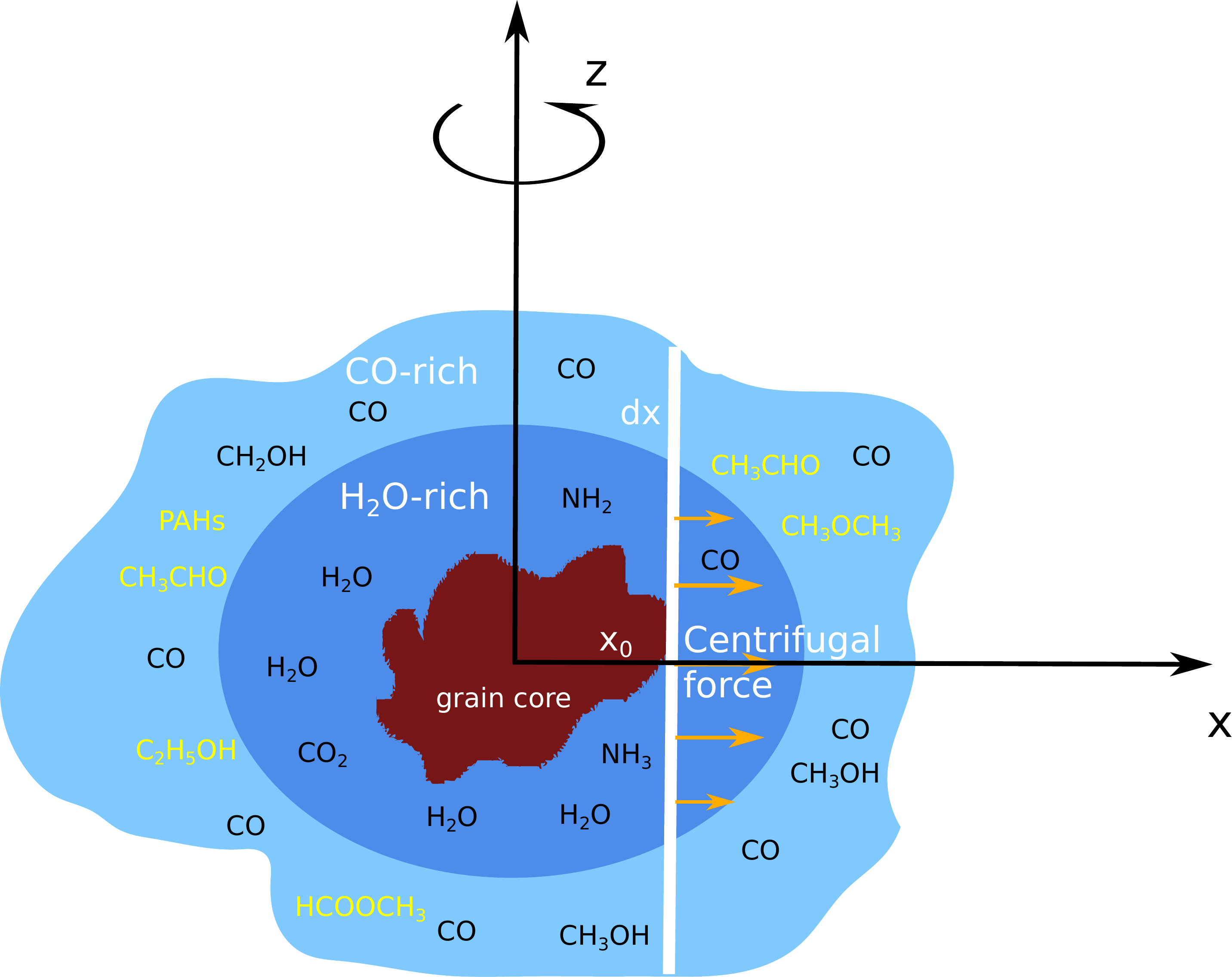}
    \caption{A schematic illustration of a rapidly spinning 
    core-mantle grain of irregular shape, comprising an icy water-rich (blue) 
    and CO-rich (orange) mantle layers. 
    The core is assumed to be compact silicate material, 
    and complex organic molecules are formed in the ice mantle of the core. 
    Centrifugal force field on a slab $dx$ is illustrated, which acts 
    to pull off the ice mantle from the grain core at sufficiently fast rotation.}
    \label{fig:model}
    \includegraphics[width=0.5\textwidth]{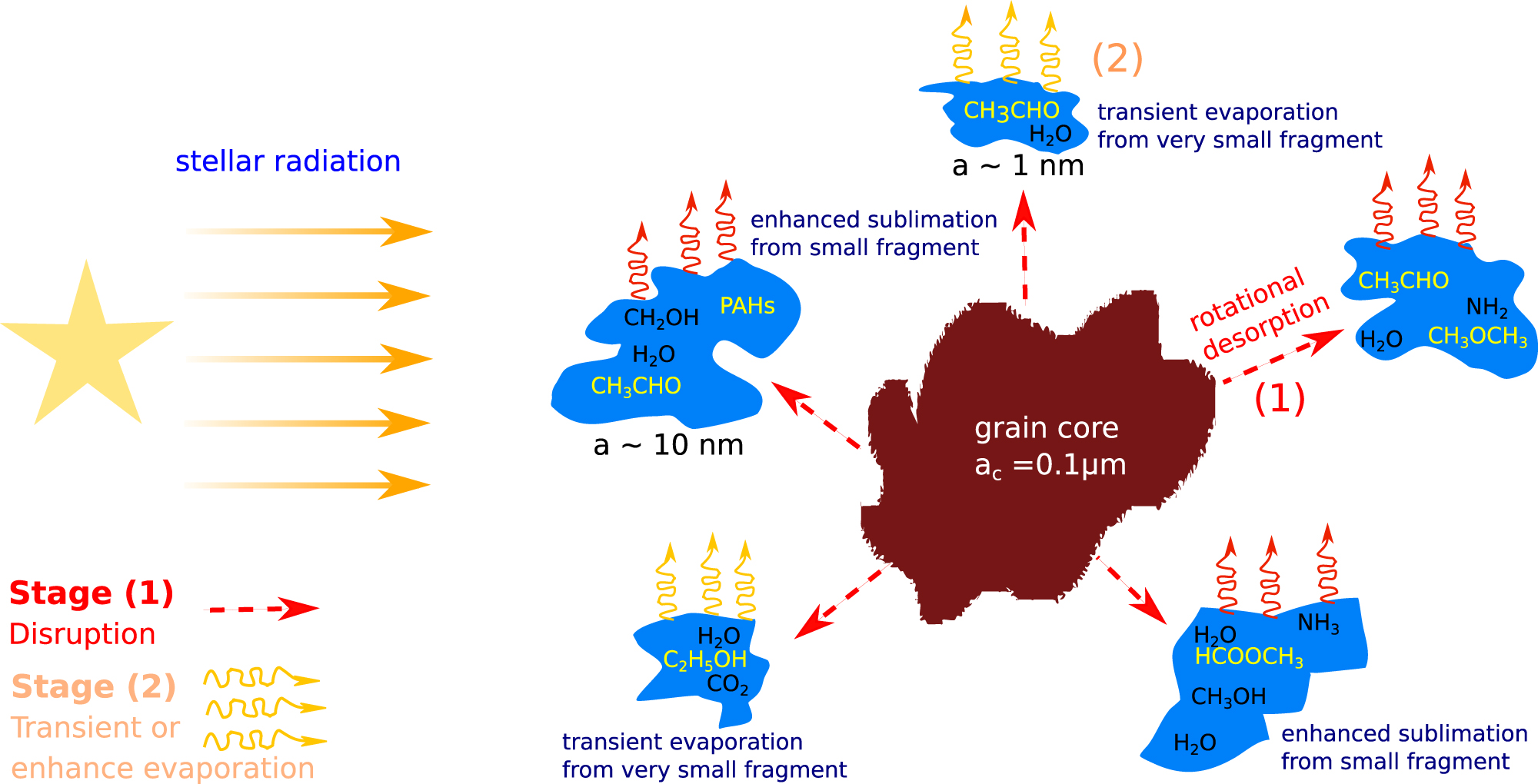}
    \caption{Schematic illustration of rotational desorption process of COMs from icy grain mantles comprising two stages: (1) disruption of icy mantes into small fragments by RATD, and (2) rapid evaporation of COMs due to thermal spikes for very small fragments or increased sublimation for larger fragments. From \cite{2020ApJ...891...38H}.}
    \label{fig:schematic}
\end{figure}
We consider an irregular grain shape with an amorphous silicate core covered by a double-ice mantle, as illustrated in Figure \ref{fig:model}. The size of the core is denoted as $a_{\rm c}$, and the average thickness of the mantle is $\Delta a_{\rm m}$. Hence, the effective radius of the grain model is $a \approx a_{\rm c} + \Delta a_{\rm}$.  

As shown in \cite{2019NatAs...3..766H}, the grain exposed to a radiation source of constant luminosity with the dimensionless radiation strength $U=u_{\rm rad}/u_{\rm ISRF}$ \footnote{$u_{\rm RAT}=\int u_{\lambda}d\lambda$ is the energy density of the radiation field. For the ISRF, U=1}(where $u_{\rm ISRF}=8.64\times 10^{-13}\erg \cm^{-3}$ is the radiation energy density of the interstellar radiation field in the solar neighborhood, \citealt{1983A&A...128..212M}) and the mean wavelength\footnote{$\bar{\lambda} = \int \lambda u_{\lambda}d\lambda /u_{\rm rad}$ depending on the temperature of the radiation source. For ISRF, $\bar{\lambda}\sim 1.2\,\mu$m} of the radiation source $\bar{\lambda}$, can be spun-up to a maximum angular velocity
\bea \label{eq:wrat_min}
    \omega_{\rm RAT} &\simeq& 9.6\times 10^{8}\gamma a^{0.7}_{-5}\bar{\lambda}^{-1.7}_{0.5} \\ \nonumber
    &&\times \left(\frac{U}{n_{1}T^{1/2}_{2}}\right)\left(\frac{1}{1+F_{\rm IR}}\right) ~~~\rm{rad\ s^{-1}}
\ena
for grains with $a \leq \frac{\bar{\lambda}}{1.8}$, and
\bea \label{eq:wrat_max}
    \omega_{\rm RAT} &\simeq& 1.78\times 10^{10}\gamma a^{-2}_{-5}\bar{\lambda}_{0.5} \\ \nonumber
    &&\times \left(\frac{U}{n_{1}T^{1/2}_{2}}\right)\left(\frac{1}{1+F_{\rm IR}}\right) ~~~\rm{rad\ s^{-1}}
\ena
for $a> \frac{\bar{\lambda}}{1.8}$. Above, $a_{-5}=a/(10^{-5} \cm)$, $\bar{\lambda}_{0.5}=\bar{\lambda}/(0.5\,\mu \rm m)$, $n_{1}=n_{\H}/(10\,\cm^{-3})$, and $T_{2}=T_{\gas}/(100\K$), the anisotropy degree $\gamma=1$. We adopt the ratio of the IR damping to the collisional damping, $F_{\rm IR}$, as in \cite{2019NatAs...3..766H}.

When the rotational angular velocity is sufficiently high, the centrifugal stress induced by this rotation can exceed the binding energy of the ice mantle to the grain core ($S_{\max}$), resulting in the disruption of the ice mantle. The critical rotational velocity is
\bea \label{eq:wdisr}
    \omega_{\disr} \simeq \frac{6.3\times 10^{8}}{a_{-5}(1-x^{2}_{0}/a^{2})^{1/2}}\hat{\rho}^{-1/2}_{\rm ice}S^{1/2}_{\rm max,7}, ~~~\rm{rad\ s^{-1}}
\ena
where $\rho_{\rm ice}$ is the mass density of ice mantles with $\hat{\rho}_{\rm ice}=\rho_{\rm ice}/(1\g\cm^{-3})$, $S_{\rm max,7}=S_{\max}/(10^{7} \erg\cm^{-3})$. Furthermore, if the rotational angular velocity is increased beyond $\omega_{\disr}$, the centrifugal stress exceeds the ice tensile strength that holds the different parts of the mantle together, resulting in the disruption of the mantle into small fragments.

We can derive the critical grain size of rotational desorption by setting $\omega_{\rm RAT}$ (Equation \ref{eq:wrat_min}) $\equiv$ $\omega_{\disr}$ (Equation \ref{eq:wdisr}), which yields:
\bea
    a_{\disr} &\simeq& 0.13\gamma^{-1/1.7}\bar{\lambda}_{0.5}(S_{\rm max,7}/\hat{\rho}_{\rm ice})^{1/3.4} \\ \nonumber
    &&\times (1+F_{\rm IR})^{1/1.7}\left(\frac{n_{1}T^{1/2}_{2}}{U}\right)^{1/1.7} ~~~\rm{\mu m.} 
\ena

The maximum size of grains that can still be disrupted is derived from Equation \ref{eq:wrat_max} and Equation \ref{eq:wdisr} as
\bea
    a_{\disr,\max} &\simeq& 2.9\gamma\bar{\lambda}_{0.5}\hat{\rho}_{\rm ice}^{1/2}S^{-1/2}_{\rm max,7} \\ \nonumber
    &&\times \left(\frac{U}{n_{1}T^{1/2}_{2}}\right)\left(\frac{1}{1+F_{\rm IR}}\right) ~~~\rm{\mu m.}
\ena

One can see that the values of $a_{\disr}$ and $a_{\disr,\max}$ depend on the radiation strength $U$ and gas density. To relate $U$ with an observable quantity that can be inferred from observations, we use a simple relationship between dust temperature and the radiation field $T_{\d}\simeq 16.4a^{-1/15}_{-5}U^{1/6} (\K)$, assuming silicate grains (\citealt{2011piim.book.....D}).

Exposed to the same strong radiation as the original large grains, the small ice fragments ($a\lesssim$ 1nm) are rapidly evaporated by transient heating, which is followed by rapid thermal desorption of COMs due to high temperatures. For larger ice fragments, the temperature increases from the radiation is higher than that of the original large grain. It therefore increases the sublimation rate of COMs from these fragments significantly. The schematic of this desorption mechanism is shown in Figure \ref{fig:schematic} (also see \citealt{2020ApJ...891...38H} for more detailed demonstrations). 

Let $f$ be the fraction of COMs formed in the ice mantle to the total ice mass, the volume mass density of COMs released to the gas phase owing to the rotational desorption of the grain-size distribution $dn/da$ grains is
\bea
    \rho(\COMs) = \int_{a_{\disr}}^{a_{\disr,\max}} f\times \frac{4}{3}\rho_{\rm ice}(a-a_{c})^{3} \left(\frac{dn}{da}\right) da ~~~\rm{g\ cm^{-3}}.
\ena
In this work, we use the grain size distribution $dn/da=An_{\H}a^{-3.5}$ with $A$ the normalization constant (\citealt{1977ApJ...217..425M}, hereafter MRN distribution). The adoption of the MRN distribution is plausible for large grains. Then, the abundance of COMs returned to the gas phase is
\bea \label{eq:xCOMs}
    x_{\rm model}(\COMs) = \frac{n(\COMs)}{n_{\H}} = \frac{\rho (\COMs)}{m(\COMs)n_{\H}}.
\ena

The parameters of our model include the radiation field (or dust temperature), gas density, and gas temperature. To reduce the number of degrees of freedom, we adopt the constraint of some parameters from observations. The uncertainties due to these choices are discussed in Section \ref{sec:uncertainty}.

\begin{figure}
    \centering
    \includegraphics[width=0.46\textwidth]{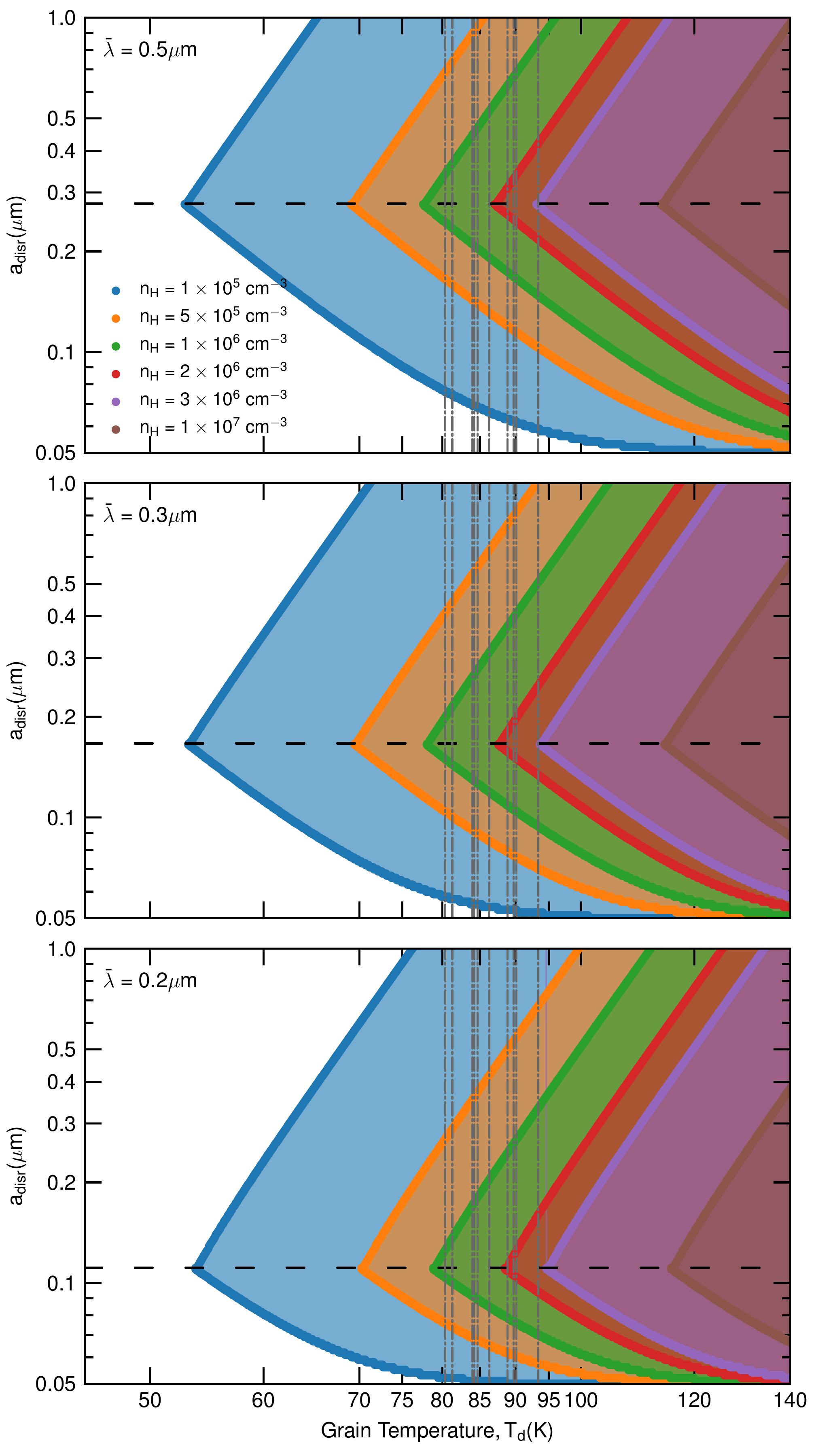}
    \caption{Range of desorption sizes of ice mantles, constrained by $a_{\disr}$ (lower boundary) and $a_{\disr,\max}$ (upper boundary), as a function of the grain temperature for the different gas densities for $\bar{\lambda} = 0.5\,\mu$m (top panel), $\bar{\lambda}=0.3\,\mu$m (middle panel), and $\bar{\lambda}=0.2\,\mu$m (bottom panel) 
    assuming a fixed core radius $a_{\rm c}=0.05\,\mu$m and the varying mantle thickness. 
    The horizontal dashed lines denote the transition size $a_{\rm trans} = \bar{\lambda}/1.8$. 
    Shaded regions mark the range of grain sizes disrupted by RATD.}
    \label{fig:adisr}
\end{figure}

\begin{table*}
\centering
\caption{The best parameter space, ($n_{\H}, f$) for some COMs with very high binding energy at three locations: MF (methyl formate) peak, EG (ethylene glycol) peak, and ET (ethanol) peak. The results are given with $S_{\rm max}=10^{7} \erg \cm^{-3}$ and $T_{\d}$ the mean value extracted with the beam size as shown in Figure \ref{fig:Td_map}. $x_{\rm obs}(\COMs)\simeq N_{\rm obs}(\COMs)/N_{\H}$ where N(COMs) and N$_{\H}$ are adopted from \cite{2018A&A...620L...6T} with N(H$_{2}$) = 0.5N$_{\H}$ assumed.}
\label{tab:COMs_compare}
\begin{tabular}{c ccc|ccc|ccc}
\toprule
\hline
\multicolumn{1}{c}{}     & \multicolumn{3}{c}{MF peak}& \multicolumn{3}{c}{EG peak} & \multicolumn{3}{c}{ET peak} \\
\multicolumn{1}{c}{}     & \multicolumn{3}{c}{(05$^{\rm h}$35$^{\rm m}$14.102$^{\rm s}$, -05$^{o}$22$'$36.84$''$)} 
       & \multicolumn{3}{c}{(05$^{\rm h}$35$^{\rm m}$14.474$^{\rm s}$, -05$^{o}$22$'$33.12$''$)} 
       & \multicolumn{3}{c}{(05$^{\rm h}$35$^{\rm m}$14.440$^{\rm s}$, -05$^{o}$22$'$34.74$''$)} \\
\multicolumn{1}{c}{}     & \multicolumn{3}{c}{$T_{\d}(\rm K)\sim$ 89.71} 
       & \multicolumn{3}{c}{$T_{\d}(\rm K)\sim$ 84.65} 
       & \multicolumn{3}{c}{$T_{\d}(\rm K)\sim$ 86.26} \\
\hline
\multicolumn{1}{c}{Observations} &  \multicolumn{9}{c}{OHCH$_{2}$CH$_{2}$OH, $E_{\rm b}=10200 \K$}\\
\hline
\multicolumn{1}{c|}{$N_{\rm obs}(\times 10^{15} \cm^{-2})$} & &$\sim$ 0.3               & & & $\sim$5.3 & & & $\sim$3.2 &  \\
\multicolumn{1}{c|}{$x_{\rm obs}(\COMs)$}                   & &$\sim 1.83\times 10^{-10}$ & & & $\sim 1.1\times 10^{-9}$  & & & $\sim 8.9\times 10^{-10}$ &  \\
\hline
\multicolumn{1}{c|}{Model} & $\bar{\lambda}$ = $0.5\,\mu$m & $0.3\,\mu$m& $0.2\,\mu$m& $0.5\,\mu$m& $0.3\,\mu$m& $0.2\,\mu$m& $0.5\,\mu$m& $0.3\,\mu$m& $0.2\,\mu$m \\
\hline
\multicolumn{1}{c|}{$n_{\H}(\times 10^{6} \cm^{-3})$}  & $\simeq$1.9 & $\simeq$ 1.3 & $\simeq$ 1.5 & $\simeq$1.3& $\simeq$ 1.4 & $\simeq$1.4 & $\simeq$1.8 & $\simeq$1.7 & $\simeq$1.5  \\
\multicolumn{1}{c|}{$f(\%)$}             & $\simeq$ 0.005  & $\simeq$0.003        & $\simeq$0.012                 & $\simeq$0.027& $\simeq$0.094 & $\simeq$0.4& $\simeq$0.13 & $\simeq$0.17 & $\simeq$0.2\\
\hline
{Observations}     & \multicolumn{9}{c}{CH$_{3}$OCH$_{2}$OH, $E_{\rm b}=7580 \K$} \\ 
\hline
\multicolumn{1}{c|}{$N_{\rm obs}(\times 10^{15} \cm^{-2})$} & &$\sim$ 15.0               & & & $\sim$120.0 & & & $\sim$200.0 &  \\
\multicolumn{1}{c|}{$x_{\rm obs}(\COMs)$}                   & &$\sim 9.1\times 10^{-9}$ & & & $\sim 2.5\times 10^{-8}$  & & & $\sim 5.6\times 10^{-8}$ &  \\
\hline
\multicolumn{1}{c|}{Model} & $\bar{\lambda}$ = $0.5\,\mu$m & $0.3\,\mu$m& $0.2\,\mu$m& $0.5\,\mu$m& $0.3\,\mu$m& $0.2\,\mu$m& $0.5\,\mu$m& $0.3\,\mu$m& $0.2\,\mu$m \\
\hline
\multicolumn{1}{c|}{$n_{\H}(\times 10^{6} \cm^{-3})$}  & $\simeq$2.0 & $\simeq$ 1.6 & $\simeq$ 1.6     & $\simeq$1.4& $\simeq$ 1.1 & $\simeq$0.9     & $\simeq$1.1 & $\simeq$0.8 & $\simeq$0.5  \\
\multicolumn{1}{c|}{$f(\%)$} & $\simeq$ 0.34 & $\simeq$0.30 & $\simeq$0.77    & $\simeq$0.90& $\simeq$0.74 & $\simeq$0.99    & $\simeq$0.62 & $\simeq$0.65 & $\simeq$0.57\\
\hline
{Observations}     & \multicolumn{9}{c}{C$_{2}$H$_{5}$OCHO, $E_{\rm b}=6250 \K$} \\
\hline
\multicolumn{1}{c|}{$N_{\rm obs}(\times 10^{15} \cm^{-2})$} & &$\sim$ 3.0               & & & $\sim$2.0 & & & $\sim$4.0 &  \\
\multicolumn{1}{c|}{$x_{\rm obs}(\COMs)$}                   & &$\sim 1.83\times 10^{-9}$ & & & $\sim 4.2\times 10^{-10}$  & & & $\sim 1.1\times 10^{-9}$ &  \\
\hline
\multicolumn{1}{c|}{Model} & $\bar{\lambda}$ = $0.5\,\mu$m & $0.3\,\mu$m& $0.2\,\mu$m& $0.5\,\mu$m& $0.3\,\mu$m& $0.2\,\mu$m& $0.5\,\mu$m& $0.3\,\mu$m& $0.2\,\mu$m \\
\hline
\multicolumn{1}{c|}{$n_{\H}(\times 10^{6} \cm^{-3})$}  & $\simeq$0.9 & $\simeq$ 1.7 & $\simeq$ 1.2 & $\simeq$1.4& $\simeq$ 1.6 & $\simeq$1.3 & $\simeq$1.6 & $\simeq$1.6 & $\simeq$1.6  \\
\multicolumn{1}{c|}{$f(\%)$}             & $\simeq$ 0.01  & $\simeq$0.09        & $\simeq$0.07   & $\simeq$0.02& $\simeq$0.42 & $\simeq$0.09& $\simeq$0.05 & $\simeq$0.13 & $\simeq$0.63\\
\hline
\end{tabular}
\end{table*}

\subsection{Numerical Results} \label{sec:numerical_results}
The local gas number density is constrained by numerous models such as photodissociation region, shocks, line fittings (e.g., \citealt{1987ApJ...315..621B}; \citealt{2007ApJ...671..536C}; \citealt{2016ApJ...829...15M}; \citealt{2018ApJ...856....9R}; \citealt{2018MNRAS.473.1472T}; \citealt{2020arXiv200302459T}), which indicate $n_{\H} > 10^{5}\,\cm^{-3}$ to a possible upper limit of order of $10^{7}\,\cm^{-3}$ (\citealt{2014ApJ...787..112C}). In such dense media, the grain size is expected to be large due to coagulation, such that we consider the maximum grain size up to $1\,\mu$m (the larger size will be discussed later). Orion BN/KL is a massive star-formation factory, stars here can be either Herbig-like star (e.g., \citealt{1984Natur.311..236T}) or OB-associations have been observed to reside (e.g., \citealt{1992MNRAS.255..594S}; \citealt{2008hsf1.book..459B}; \citealt{2012A&A...547A..97A}). Therefore, we consider three representative values of $\bar{\lambda}$ of $0.5\,\mu$m ($T_{\ast} \cong 10^{4}\,\K$), $0.3\,\mu$m ($T_{\ast} \cong 2\times 10^{4}\,\K$) and $0.2\,\mu$m ($T_{\ast} \gtrsim 3\times 10^4\,\K$).

Figure \ref{fig:adisr} shows the range of desorption size of ice mantles defined by $a_{\disr}$ and $a_{\disr,\max}$ as a function of the radiation field, equivalent, the dust temperature ($T_{\d}$), assuming a grain core $a_{c}=0.05\,\mu$m and the ice thickness varied up to $\Delta a_{m}=0.05\,\mu$m. The shades indicate where the rotational disruption that takes place, and the vertical lines correspond to the mean value of $T_{\d}$ at each ALMA location (see Figure \ref{fig:Td_map}). Note that the model in which the ice mantle thickness is fixed and grain core is varied is ineffective for the medium with $n_{\H}\geq 10^{6} \cm^{-3}$ and $T_{\d}<100\,\K$ (see Figure 5 in \citealt{2020ApJ...891...38H}). Therefore, we don't consider this model in this work.   

For a given $\bar{\lambda}$, the disruption size depends on the radiation field (i.e., dust temperature) and the gas number density ($n_{\H}$). 
The decreasing $a_{\disr}$ indicates that stronger radiation can disrupt smaller grain sizes, e.g., the ice mantle starts being removed at $T_{\d}<55\K$ for $n_{\H} = 10^5\cm^{-3}$ and entirely removed at $T_{\d} \simeq 120\K$ with $\bar{\lambda}=0.5\,\mu$m (see top panel). Along with that, dust grains in the denser media require stronger radiation field to be disrupted, e.g., $T_{\d} = 80\K$ can remove the ice mantle of grains with $n_{\H}=10^6\cm^{-3}$, which is higher in comparison than $n_{\H}=10^5\cm^{-3}$. For $n_{\H}>3\times 10^6\cm^{-3}$, stronger radiation with $T_{\d}>100\K$ is required to disrupt the ice mantles. The mantle is removed much more efficiently at smaller value of $\bar{\lambda}$ due to much stronger radiative torques (see middle and bottom panels).

As mentioned at the beginning, we test this mechanism by calculating the abundance of some COMs with very high binding energy, including the ethylene glycol (OHCH$_{2}$CH$_{2}$OH, E$_{\rm b}$=10200 K), the methoxymethanol (CH$_{3}$OCH$_{2}$OH, E$_{\rm b}$=7580 K) and the ethyl formate (C$_{2}$H$_{5}$OCHO, E$_{\rm b}$=6250 K) at three locations (the MF peak, the EG peak, and the ET peak) given by Equation \ref{eq:xCOMs} by varying the gas number density $n_{\H}$ in [$10^{5}-10^{7}\cm^{-3}$] intervals, and the fraction of COMs on the mantle, $f$, in [0.00001$\%$ - 1$\%$] intervals, as two independent free parameters. For each value of ($n_{\H}, f$), we calculate the relative error to the observed abundance as $RE =|x_{\rm mod}(\COMs) - x_{\rm obs}(\COMs)|/x_{\rm obs}(\COMs)$. 
The best ($n_{\H},f$) space shown in Table \ref{tab:COMs_compare} is derived from the minimum value of $RE$\footnote{The best parameters are strongly sensitive to the number of space sampling and thus subject to vary. In Table \ref{tab:COMs_compare}, the lengths of the $n_{\H}$ and $f$ intervals are 28 and 30000, respectively.}. Table \ref{tab:COMs_compare} indicates that the amount of these large COMs formed on the ice mantles is as small as $f < 1.0\%$. As for a comparison, the relative abundance of OHCH$_{2}$CH$_{2}$OH to water is $0.03\%$ toward the hot core in Orion BN/KL (see Table B.1 in \citealt{2016A&A...587L...4C}), while that is $0.2\%$ on the comet 67P (see Table 1 in \citealt{2015Sci...349b0689G}). One can see that rotational desorption could be a plausible mechanism to explain the existence of these large COMs.

Our calculations are obtained for a grain size of $0.1\,\mu$m. Nevertheless, observations show evidence for the existence of large grains in Orion as discussed above. In a dense medium, coagulation is expected to form large grains owing to grain-grain collisions, which could result in large accumulations of original ice grains, the so-called composite grains. As the large grains are exposed to the same strong radiation field, they are expected to be disrupted by the RATD mechanism. Their disruptions can be calculated as described in Section \ref{sec:rot_desorp}, but with the tensile strength of the composite grains ($S_{\max}$ from $10^{4}-10^{6} \erg \cm^{-3}$, \citealt{2018MNRAS.479.1273G}). The detailed calculations are implemented in Section 5.3 in \cite{2020ApJ...891...38H}, which showed that the large aggregate grains are rapidly disrupted into smaller grains with ice mantles within $30\,\K <T_{\d}<100\,\K$ for $n_{\H}<10^{8}\,\cm^{-3}$. Then, COMs could be desorbed by the rotational desorption mechanism applied for the smaller grains within ice mantles. 
For denser medium of $n_{\H}>10^{8}\,\cm^{-3}$, the RATD is ineffective within $T_{\d}<100\,\K$, and large grains are expected to survive (see Figure 10 in \citealt{2020ApJ...891...38H}). 

\subsection{Comparing the rotational desorption to sublimation and photodesorption mechanisms}
\cite{2020ApJ...891...38H} demonstrated that rotational desorption is more effective than the sublimation for the same size of grain core, as well as than the photodesorption mechanism.  
In this section, we recall the timescales of those mechanisms for the comparison.

The rotational timescale is the time necessarily for a grain of size $a$ reaches $\omega_{\rm RAT}$. In a strong radiation field, the disruption timescale is
\bea \label{eq:tdisr}
    t_{\disr} \simeq 0.6(\gamma U_{5})^{-1}\bar{\lambda}^{1.7}_{0.5}\hat{\rho}^{1/2}_{\rm ice}S^{1/2}_{\rm max,7}a^{-0.7}_{-5} ~~~\rm{yr}
\ena
for $a_{\disr} < a< \bar{\lambda}/1.8$, and
\bea
    t_{\disr} \simeq 0.04(\gamma U_{5})^{-1}\bar{\lambda}^{-1}_{0.5}\hat{\rho}^{1/2}_{\rm ice}S^{1/2}_{\rm max,7}a^{2}_{-5} ~~~\rm{yr}
\ena
for $\bar{\lambda}/1.8 < a<a_{\disr,\max}$

The timescale of the thermal sublimation is the time it takes for a grain core to sublimate the ice mantle of size $\Delta a_{\rm}$ as
\bea \label{eq:tsub}
    t_{\rm sub} \sim 1.5\times 10^{3}\left(\frac{\Delta a_{\rm m}}{500\AA}\right)\exp{\left(\frac{E_{b}}{4800\,\K}\frac{100\,\K}{T_{\d}}\right)} ~~~ \rm{yr}
\ena
where $E_{\rm b}$ is the binding energy of the ice mantle to the grain core.

The timescale of the photodesorption is the time required to remove the ice mantle of size $\Delta a$ by incident UV photons
\bea \label{eq:tpd}
    t_{\rm pd} \simeq 2.2\left(\frac{\Delta a_{\rm m}}{500\AA}\right)\left(\frac{10^{-3}}{Y_{\rm pd}}\right)\left(\frac{10^{5}}{G_{0}}\right) ~~~\rm{yr}
\ena
where $Y_{\rm pd}$ is the photodesorption yield (i.e., the fraction of molecules ejected over the total number of incident UV photons), $G_{0}$ is the strength of the UV photons relative to the standard interstellar radiation field. $Y_{\rm pd}$ does vary on orders of magnitude from one molecule to the other (\citealt{2019ECS.....3.1135F}).
\begin{figure}
    \centering
    \includegraphics[width=0.45\textwidth]{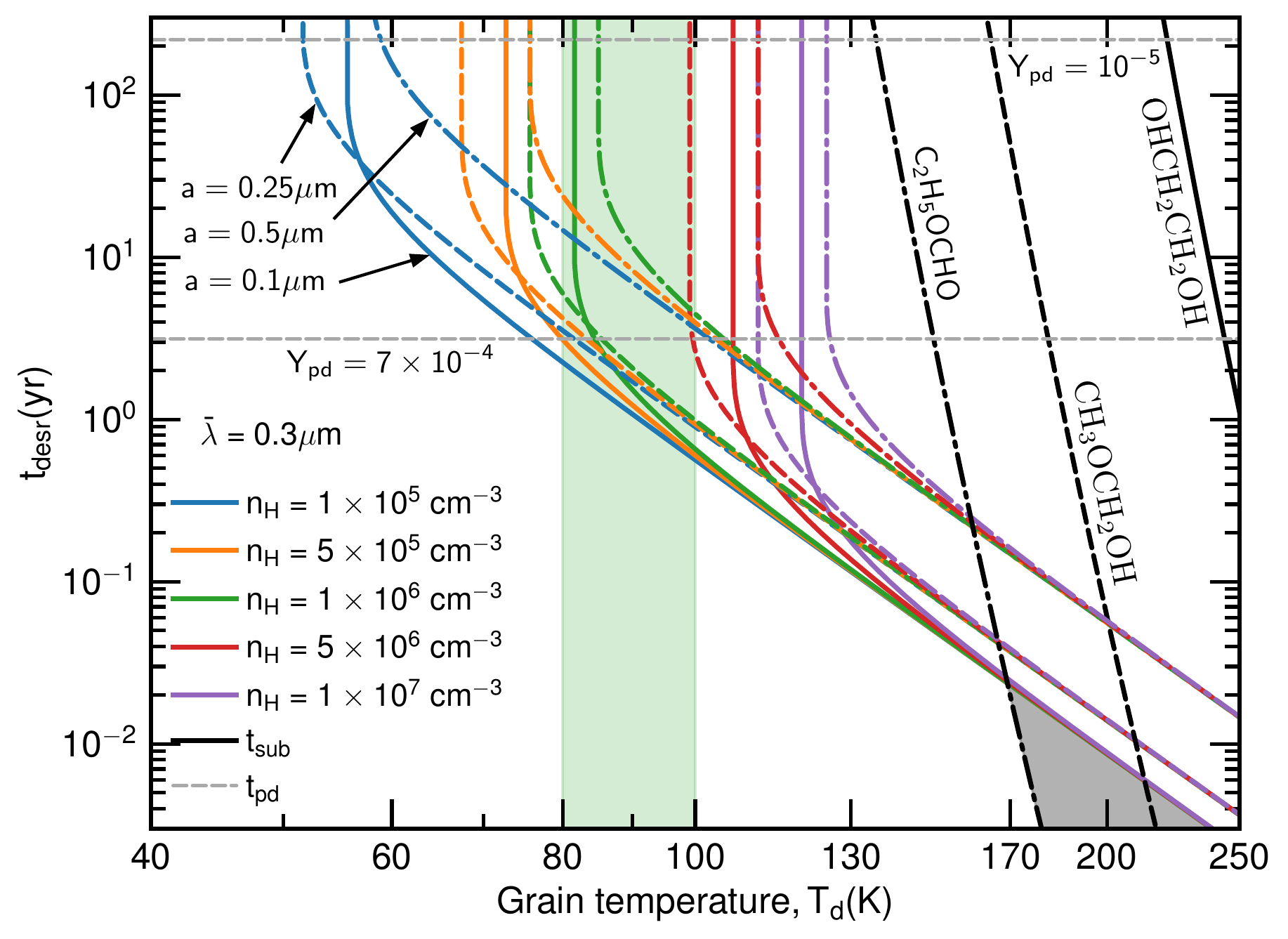}
    \caption{Timescales of the rotational desorption, sublimation and the UV photodesorption. The green area is where we are interested in, and the grey area is where the sublimation is effective.}
    \label{fig:timescales}
\end{figure}
The timescales are shown in Figure \ref{fig:timescales}. One can see that the timescale of the rotational desorption is much shorter than that of the sublimation at $T_{\d}\leq 100\,\K$, while the thermal sublimation of OHCH$_{2}$CH$_{2}$OH is dominant over at $T_{\d}\geq 300\,\K$. This value is lower for other COMs with lower binding energy, e.g., $T_{\d}\geq 170\,\K$ for C$_{2}$H$_{5}$OCHO.
If we assume $G_{0}=U_{0}=10^{5}$, and adopt $Y_{\rm pd}\simeq 7\times 10^{-4}$ as for H$_{2}$CO (i.e., 4-10 $\times 10^{-4}$, \citealt{2019ECS.....3.1135F}), the rotational desorption mechanism is shorter than the UV photodesorption at $n_{\H}<5\times 10^{6}\,\cm^{-3}$ with $80\,\K \leq T_{\d}\leq 100\,\K$. However, \cite{2016ApJ...817L..12B} experimentally showed that the photodesorption yield is an order of $10^{-5}$ for pure methanol ices, and even lower than $10^{-6}$ when it mixes with CO molecules in the ice. In this case, the rotational desorption is dominant over the UV photodesorption. Furthermore, as discussed in \cite{2020ApJ...891...38H} the photodesorption is effective with the FUV-photons, while the rotational desorption can work for a broader range of wavelength since the RATs depends on the ratio of wavelength to grain size. The penetration length of the FUV photons is shorter than longer photon wavelength (e.g., optical), so that the rotational desorption is expected to be effective in more extended regions. 
Nevertheless, Orion BN/KL is a complex peculiar star-forming region, the outflow and/or the explosion also favour COMs releasing from the ice mantles to the gas phase through the non-thermal processes (e.g., non-thermal sputtering and grain collision as discussed by \citealt{2018A&A...620L...6T}) and/or the the impact of the bullet ejected from the explosion (see \citealt{2017A&A...604L...2F} and \citealt{2017ApJ...843...83W}).

\begin{figure*}
    \centering
    \includegraphics[width=0.45\textwidth]{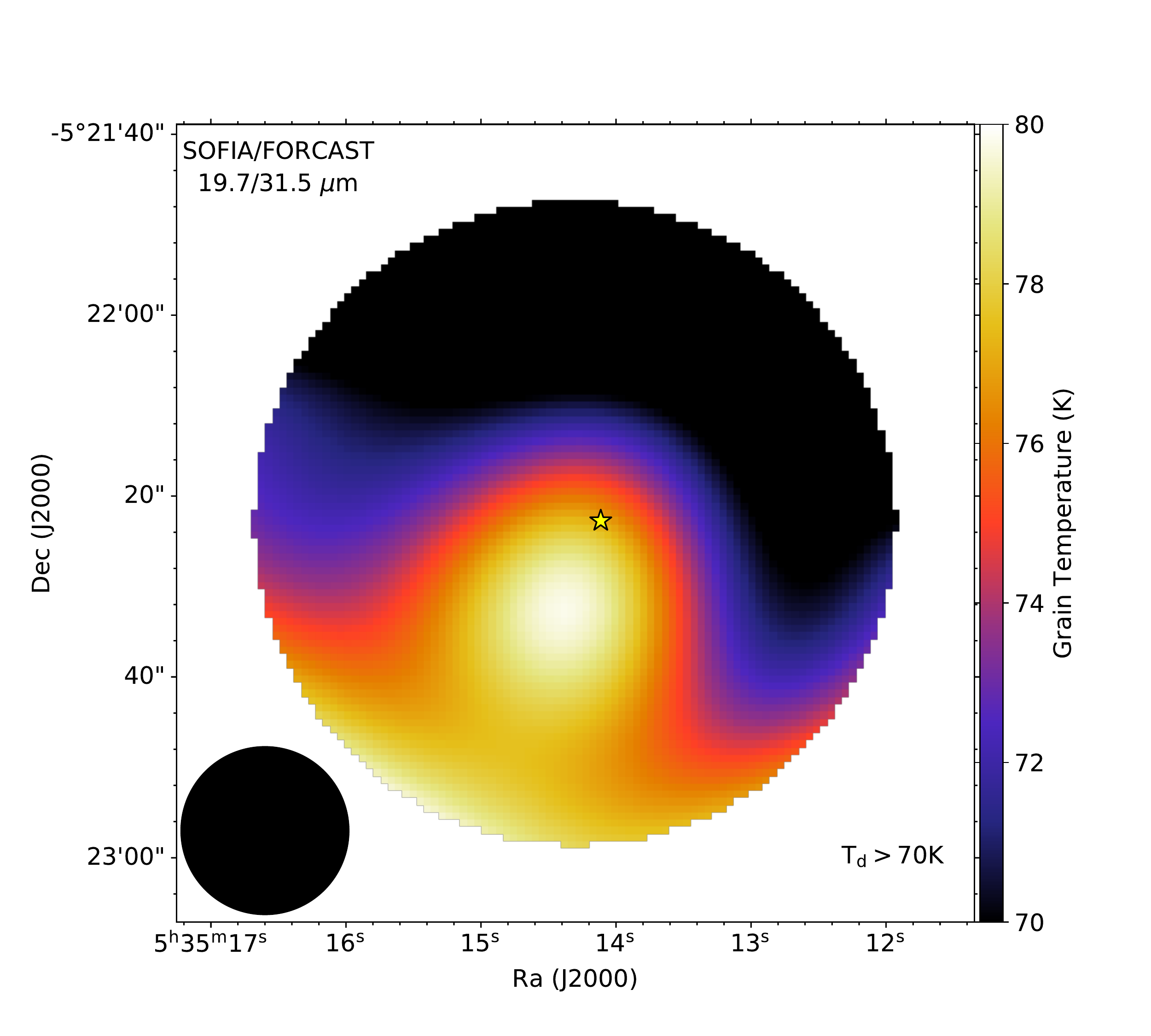}
    \includegraphics[width=0.45\textwidth]{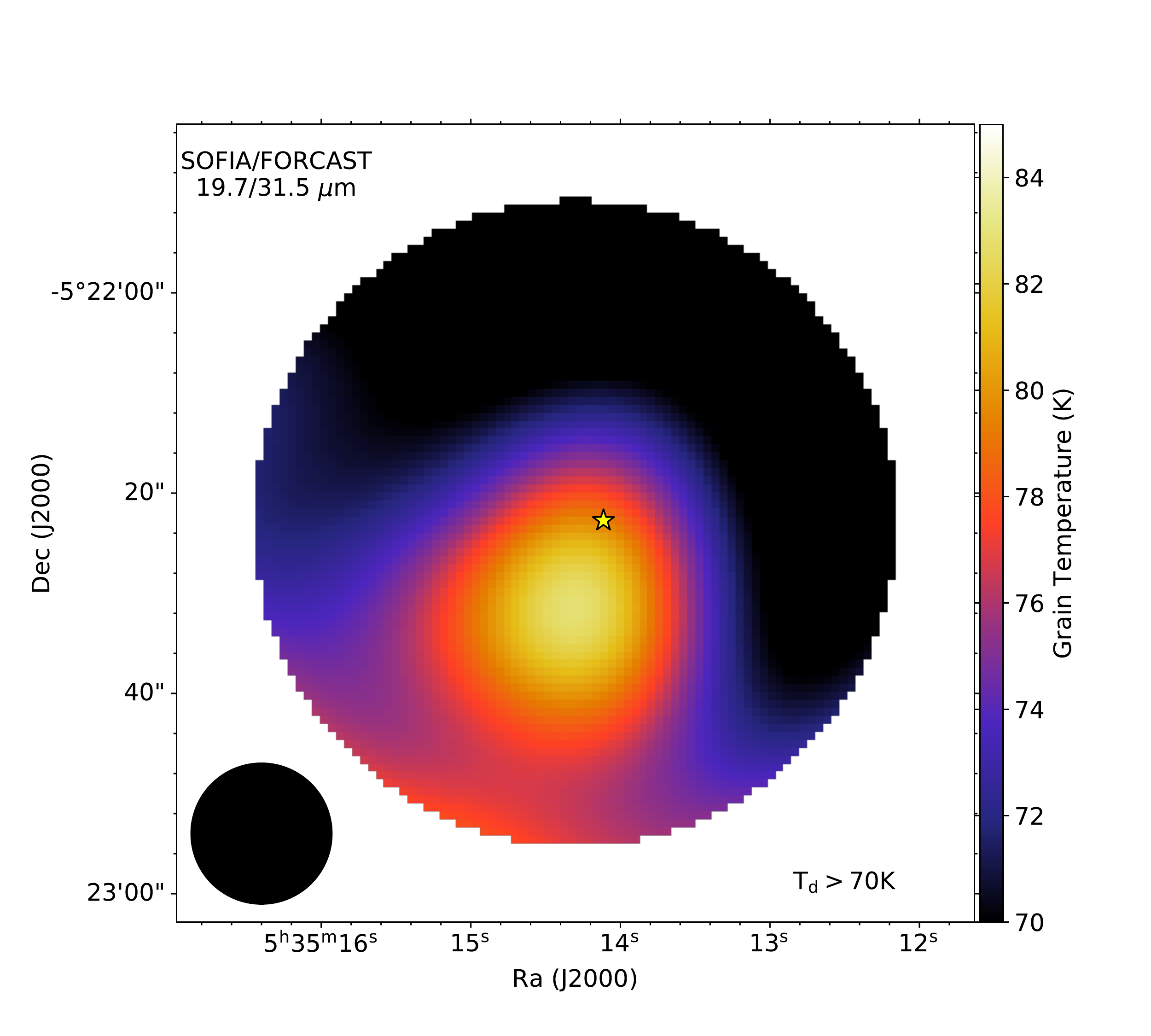}
        \caption{Map of the dust temperature with $T_{\d}>70\,\K$ smoothed to 18.7$''$ (\textit{left panel}), and to 14.2$''$ (\textit{right panel}).}
        \label{fig:Td_smooth}
        
    \includegraphics[width=0.45\textwidth]{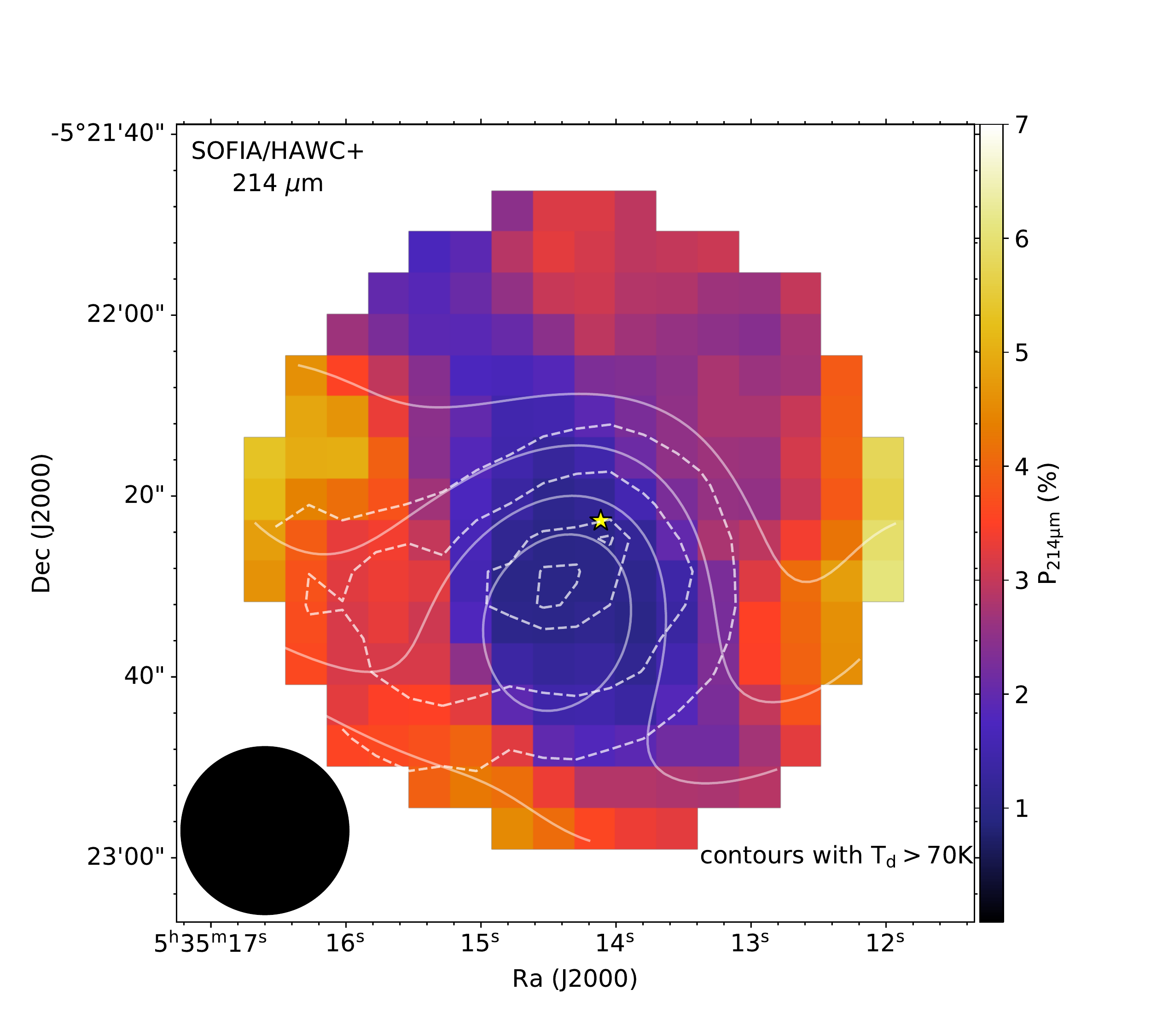}
    \includegraphics[width=0.45\textwidth]{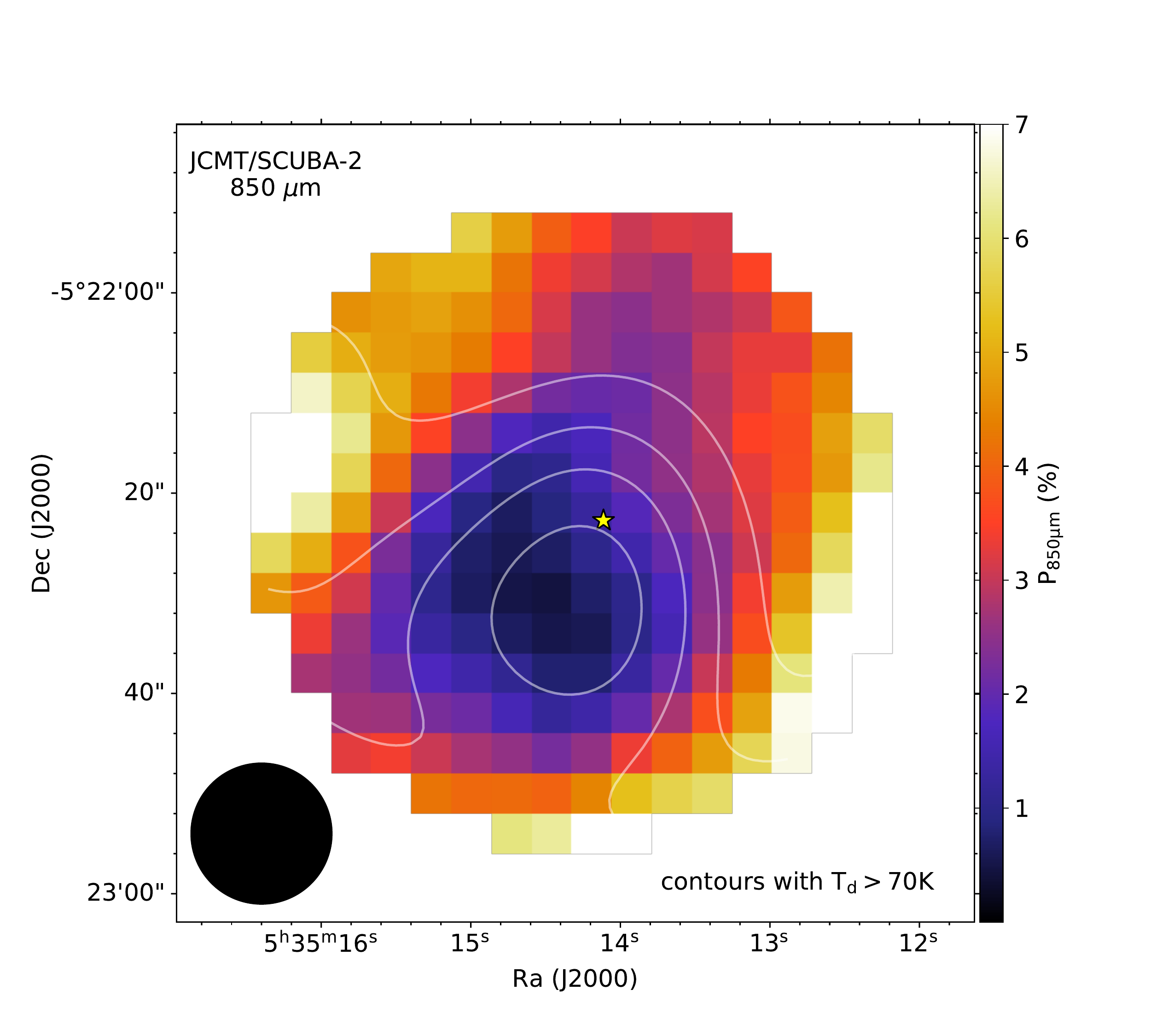}
        \caption{Polarimetry observations toward Orion BN/KL. \textit{Left panel:} 18.7$''$ Map of dust polarization degree observed by SOFIA/HAWC+ at 214$\,\mu$m. \textit{Right panel:} 14.2$''$ Map of dust polarization degree observed by JCMT/SCUBA-2 at 850$\,\mu$m. The solid contours indicate the dust temperature maps in Figure \ref{fig:Td_smooth}. The dashed contours show the 18.7$''$ map of the dust temperature from \cite{2019ApJ...872..187C}.}
        \label{fig:pol_contour}
\end{figure*}
\section{Correlation of dust polarization and COMs with dust temperature} \label{sec:polarization_COMs}
In Section \ref{sec:rot_desorp}, we showed that RATD could reproduce the detection of COMs in the Orion BN/KL. In this section, we seek for additional evidence of RATD by using the variation of dust polarization degree with the dust temperature.

In the framework of RAT theory, the polarization degree of dust emission is expected to increase with the radiation intensity (\citealt{2020ApJ...896...44L}). However, radiative torques of the very intense radiation field disrupt large grains into smaller species (\citealt{2019NatAs...3..766H}). Consequently, the depletion of large grains reduces the polarization degree, as demonstrated in \cite{2020ApJ...896...44L} and \cite{2020arXiv200710621T}. As RATD effectively disrupts the large grains that dominate long-wavelength emission, we expect a decrease of the polarization in far-infrared/sub-millimeter when the radiation intensity or grain temperature increases.

\begin{figure*}
    \centering
    \includegraphics[width=0.9\textwidth]{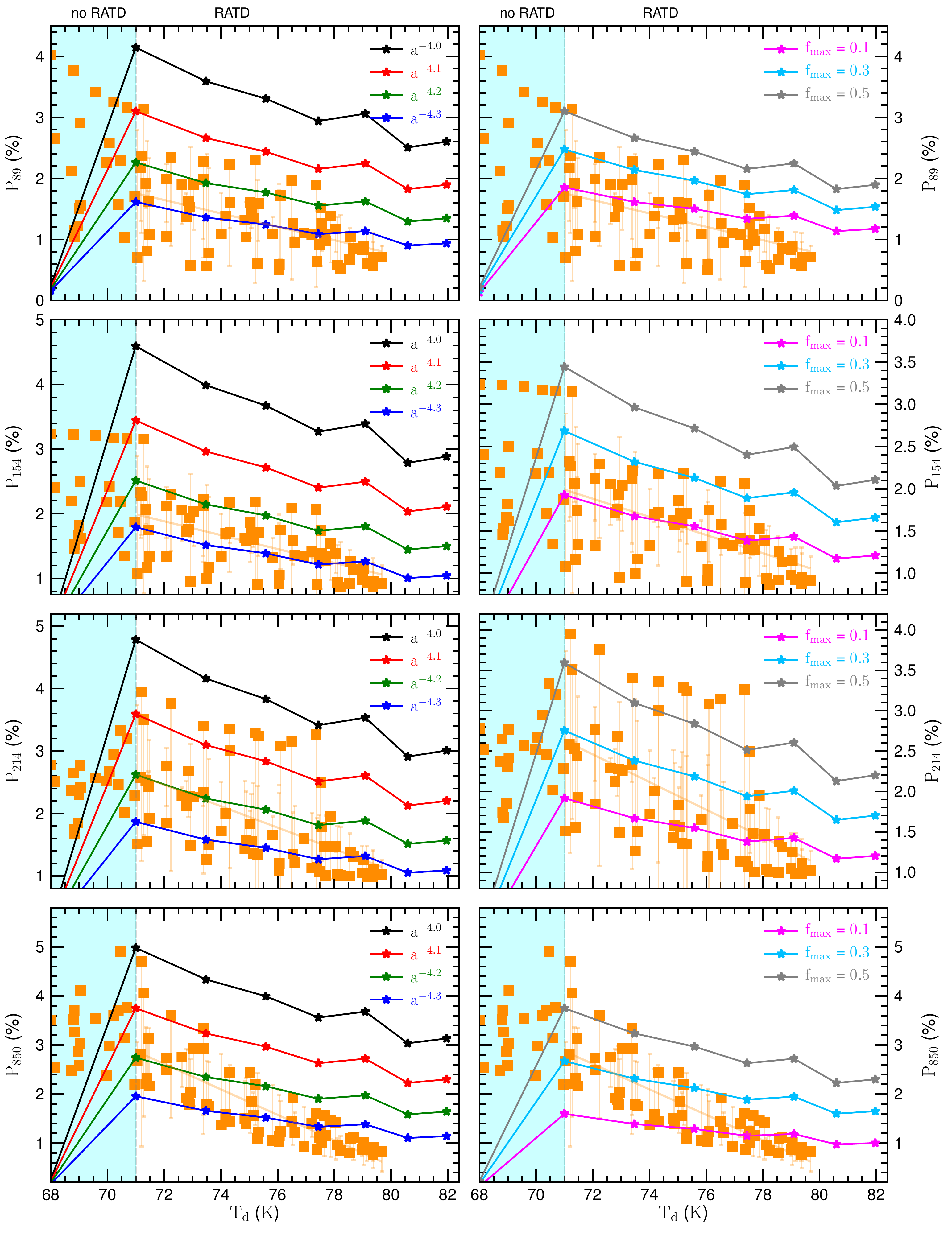}
    \caption{Variation of dust polarization degree with dust temperature at four wavelengths observed by SOFIA/HAWC+ and JCMT/SCUBA-2, from top to bottom: $\lambda=89\,\mu{\rm m}$, $\lambda=154\,\mu{\rm m}$, $\lambda=214\mu{\rm m}$ and $\lambda=850\mu{\rm m}$. Orange points are the observational data. Faint solid orange line, whose errors stand for the residual of each data point, is the weighted-fitting to a linear function. Color lines are the polarization models of thermal dust emissions. Left panel: models for various values of the power index of the grain-size distribution $\beta$ assuming $f_{\max}=1.0$ for silicate grains and $f_{\max}=0.5$ for graphite grains. Right panel: models with varying $f_{\max}$ of graphite grains and fixed $\beta=4.1$. Where RATD does not occur (cyan shades), the polarization degree increases with dust temperature. The polarization degree decreases as dust temperature increases when RATD occurs (white regions).}
    \label{fig:model_one_wave}
\end{figure*}

\subsection{Modeling thermal dust polarization}
We model polarized thermal emission from aligned dust grains with RATD as described in \cite{2020ApJ...896...44L}. enThe gas density and radiation strength are estimated as in Figure \ref{fig:Td_map} and Table \ref{tab:COMs_compare}. The gas density is 5$\times 10^6\,\cm^{-3}$ at a center, where $U>20000$ (corresponding to the region where $T_{\d}$ is the highest, the center in Figure \ref{fig:Td_map}), $n_{\gas}$=3$\times 10^6\,\cm^{-3}$ at $77\K<T_{\d}<85\K$, $n_{\gas}$=$10^6\,\cm^{-3}$ at $71\K<T_{\d}<77\K$, and $n_{\gas}$=1$\times 10^5\,\cm^{-3}$ at $T_{\d}<71\K$.
The radiation strength is estimated from the dust temperature as $U=(T_{d}/16.4\K)^{6}$ (see also Section \ref{sec:rot_desorp}). We assume a power-law grain size distribution, $dn/da\propto a^{-\beta}$ with the lower and upper cutoff grain sizes of $a_{\rm min}=3.5\times 10^{-4}\,\mu{\rm m}$ and $a_{\rm max}=1\,\mu{\rm m}$, respectively. From the ISM, the grain size distribution follows the MRN-distribution with power-law index, $\beta = 3.5$. However, toward intense radiation sources, one expects that RATD enhances the abundance of small grains, resulting in a steeper slope. Thus, we consider different slopes from $\beta=4.0-4.3$.

The RATD mechanism is efficient for the large grains and constrains the upper cutoff of the grain size distribution (\citealt{2019ApJ...876...13H}). The internal structure of grains is determined by the tensile strength ($S_{\rm max}$). More compact grains have higher values of $S_{\rm max}$. For instance, the compact grain has $S_{\max}\geq 10^9\erg\,\cm^{-3}$, while the composite grain has $S_{\max} \sim 10^7\erg\,\cm^{-3}$. However, the grains in molecular cloud seem to be in the composite structure, thus we use $S_{\rm max}=10^{7}\erg \cm^{-3}$ in our model (the effect of $S_{\rm max}$ is considered in \citealt{2020arXiv200710621T}). To model the polarization from dust heated by a hot source in a molecular cloud, we adopt a mixed dust model consisting of two separate populations of amorphous silicate grains and carbonaceous (graphite) grains (see \citealt{2007ApJ...657..810D}). We assume a prolate spheroidal shape with axial ratio $r = 1/3$ for both silicate and carbonaceous grains. The size-dependence alignment degree is described by a function $f_{\rm align}(a)=f_{\rm min}+\left(f_{\max}-f_{\rm min}\right)\left[1-\exp\left(0.5a/a_{\rm align}\right)^3\right]$. Here, we assume that silicate grains can be perfectly aligned with $f_{\max}=1$, whereas graphite grains are partially aligned, i.e., $f_{\max}<1$. The minimum alignment degree $f_{\rm min}$ is fixed to $0.001$. 
The interpretation to observations is discussed in Section \ref{sec:214um}.

\subsection{Observed Correlation versus Theoretical Modeling}\label{sec:214um}
Figure \ref{fig:Td_smooth} shows the maps of the dust temperature smoothed to $18.7''$ (corresponding to 214$\,\mu$m SOFIA/HAWC+ beam size, left panel) and to $14.2''$ (corresponding to 850$\,\mu$m JCMT/SCUBA-2, right panel) in Orion BN/KL. These maps indicate that $T_{d}\gtrsim 71\K$ in the South-East direction from the BN/KL (where COMs are detected) and is hotter than the North-West direction. Figure \ref{fig:pol_contour} shows the spatial correlation of the polarization degree maps of 214$\,\mu$m SOFIA/HAWC+ (left panel) and 850$\,\mu$m JCMT/SCUBA-2 (right panel) in the same region overlaid with the temperature in Figure \ref{fig:Td_smooth} (white contours) within the same beam size. Additionally, we also overlaid the $18.7''$ dust map constructed from a combination data (see Section \ref{sec:obs_info}) on the 214$\,\mu$m SOFIA/HAWC+ as the white dashed contours for a comparison. According to Figure \ref{fig:pol_contour}, the peak of the dust temperature coincides with the minimum of the polarization degree, in which COMs are detected.

Figure \ref{fig:model_one_wave} shows the scatter plots (orange points) of the polarization degree at 89$\,\mu$m, 154$\,\mu$m, 214$\,\mu$m and 850$\,\mu$m with dust temperature from top to bottom. For comparison, we simply smooth all the data to $18.7''$ which is a maximum beam size in our data set. For $T_{\d}<71\K$, the polarization degree increases as increasing dust temperature (positive slope), and it is higher for longer wavelength ($P_{850\,\mu{\rm m}} > P_{214\,\mu{\rm m}} > P_{154\,\mu{\rm m}} \gtrsim P_{89\,\mu{\rm m}}$). For $T_{\d}\gtrsim 71\K$, the polarization degree decreases with dust temperature (negative slope). The weighted-fitting to a linear function indicates that the slope ($\alpha$) of the negative trend is steeper for longer wavelength, i.e., $\alpha_{850\,\mu {\rm m}}=-0.26$, $\alpha_{214\,\mu {\rm m}}=-0.17$, $\alpha_{154\,\mu {\rm m}}=-0.11$, $\alpha_{89\,\mu {\rm m}}=-0.11$. The fits are shown by the faint solid orange lines within an error bar of each data point. \cite{2019ApJ...872..187C} explained the negative slope by the purely stochastic variation of the magnetic field along the line-of-sight on the scale of the beam. Nevertheless, this mechanism could not explain the positive slope phenomena.

In Figure \ref{fig:model_one_wave}, we see the results of dust polarization from modeling at 89$\,\mu$m, 154$\,\mu$m, 214$\,\mu$m, and 850$\,\mu$m. They are represented as coloured solid lines. As both the power-law index of the grain-size distribution ($\beta$) and the degree of grain alignment ($f_{\max}$) control the polarization degree of thermal dust emission, we obtain the numerical results for different values of these parameters. The left panels are for a variation of $\beta$ with $f_{\max}$ fixed, while the right panels are for a range of $f_{\max}$ with $\beta$ fixed. The model can reproduce both the increase and then the decrease of the polarization degree with dust temperature with the transition at $71\K$. The reason is as follows. For $n_{\H}>5\times 10^{5}\,\cm^{-3}$, the RATD mechanism does not occur at $T_{\d}<71\,\K$ (see Figure \ref{fig:adisr}), thus the polarization fraction increases with increasing dust temperature as predicted by the RAT alignment theory. In addition, the polarization degree is higher for longer wavelengths owing to the predominance of large grains in Orion BN/KL region (see Section \ref{sec:obs_info}). When the temperature increases to $T_{\d}\gtrsim71\,\K$, the RATD mechanism starts to occur and large grains are disrupted into smaller fragments, decreasing the abundance of large grains. As a result, the polarization degree decreases with increasing the grain temperature. This successfully reproduces the observational data. Furthermore, since RATD works effectively for larger grains, the corresponding decline slope is steeper for longer wavelengths which are dominantly produced by large grains. It is reflected in grain size distribution. The left panels show that the size distribution of dust grains with high $T_d$ has steeper power-law index due to enhancement of small grains.

\subsection{Uncertainties of adapted physical properties} \label{sec:uncertainty}
Throughout this work, we used the values of gas temperature, gas number density, and especially dust temperature inferred from observations. These values are in fact the averaged ones and contain bias of the projection effects, which are not the local values. The different values of the gas temperature and number density could change our constraints on $n_{\H}$ and $f$ but could not affect our main conclusions. The main concern would be the dust temperature because $T_{\d}$ is the averaged dust temperature (i.e., projected in the plane of the sky) and not the local value. So, one cannot rule out the possibility that COMs have been sublimated if local dust temperature is as high as $T_{\rm sub}$. Therefore, the sublimation becomes an important desorption process. However, if there is such a case, based on the RATD mechanism principles, we can see that the rotational desorption already removes the ice-mantle before the grains are being heated to such a high temperature. Thus, the rotational desorption would have to be considered as one of the efficient mechanisms to desorb COMs. Moreover, the successful explanation of dust polarization vs. temperature using the RATD mechanism is an important constraint for rotational desorption of COMs. 

\section{Summary} \label{sec:summary}
Complex organic molecules (or COMs) are believed to form in the ice mantle of dust grains and are released to the gas by thermal sublimation ($T_{\rm d}>100\K$). However, some large COMs having very high binding energy are detected in regions where $T_{\rm d}$ is far away from the sublimation temperature. \cite{2020ApJ...891...38H} proposed a new mechanism the so-called rotational desorption that can desorb COMs at low temperatures. In this work, we have searched for observational evidence for rotational desorption of COMs induced by RATs using chemical abundance and dust polarization in the massive star-forming region Orion BN/KL. Our main findings are summarized as follows:

\begin{itemize}
\item[1] Observations toward Orion BN/KL reveal the presence of numerous COMs in the regions with dust temperatures below their sublimation threshold. This reveals that thermal sublimation cannot explain the desorption of COMs from grain surface.

    \item[2] We explain the detection of COMs by numerically modeling of rotational desorption of ice mantles from grain surfaces for Orion BN/KL. We find that in highly dense region ($n_{\H} \geq 10^{6}\,\cm^{-3}$), large ice mantle-core grains can be disrupted into tiny fragments when the grain temperature increases to $T_{\d}>80\K$.  

    \item[3] We calculate the abundance of some large COMs (OHCH$_{2}$CH$_{2}$OH, CH$_{3}$OCH$_{2}$OH and C$_{2}$H$_{5}$OCHO) with the binding energy (E$_{\rm b}=10200\K, 7580\K, 6250\K$) much larger than that of water (E$_{\rm b}=5700\K$) in three different locations (the methyl formate peak, the ethylene glycol peak and the ethanol peak) by the rotational desorption. Our mechanism can reproduce the ALMA observations (\citealt{2018A&A...620L...6T}) with the gas number density in order of 10$^{6}\,\cm^{-3}$ and the fraction of COMs ($f < 1.0\%$) formed on ice mantles. This fraction seems consistent with another ALMA constraint toward the hot core clump in Orion BN/KL (\citealt{2016A&A...587L...4C}), and the in-situ experiment on the comet 67P (COSAC aboard Rosetta's Philae lander, \citealt{2015Sci...349b0689G}).   
        
    \item[4] For regions with gas density of $n_{\H}<5\times 10^{6}\,\cm^{-3}$, the rotational desorption of large icy mantles ($a<0.5\,\mu$m) is more effective than the photodesorption within $80\,\K \leq T_{\d}<100\,\K$, or at least comparable with regards to the photodesorption yield uncertainty. The effect of the rotational desorption is expected to be more extended than the photodesorption.
    
    \item[5] We analyzed multiple-wavelength polarimetric data from SOFIA/HAWC+ (89$\,\mu$m, 154$\,\mu$m, and 214$\,\mu$m) and JCMT/SCUBA-2 (850$\,\mu$m) toward Orion BN/KL and demonstrate a correlation between dust polarization and COMs. The polarization degree tends to increase with increasing the dust temperature and then decreases when the grain temperatures exceeds $T_{\d}\sim 71\K$.
    
    \item[6] We performed numerical modeling of dust polarization by aligned grains due to RATs, taking into account rotational disruption. Our obtained results successfully reproduce the observed anti-correlation between dust polarization and temperature. This supports the rotational disruption mechanism and our explanation of desorption of COMs in terms of rotational desorption.
        
\end{itemize}
To sum up, we showed that the rotational desorption is plausible to explain the existence of large COMs with very high binding energy at low temperature region (i.e., lower than the sublimation temperature) in Orion BN/KL. This mechanism is then confirmed by the observations of the polarized thermal dust emission. Therefore, we propose that the rotational desorption should be considered as one the important mechanism to desorb COMs from the grain ice mantles to the gas-phase along with other non-thermal processes.
\\ \\
Acknowledgments: We thank the anonymous referee for helpful comments and suggestions that improved the presentation of this paper significantly. We thank Dr. James De Buizer for sharing the map of dust temperature constructed from SOFIA/FORCAST observations, Dr. Wanggi Lim, Dr. Thanh Nguyen and Dr. Lucas Grosset for useful comments. This research is based on observations made with the NASA/DLR Stratospheric Observatory for Infrared Astronomy (SOFIA). SOFIA is jointly operated by the Universities Space Research Association, Inc. (USRA), under NASA contract NNA17BF53C, and the Deutsches SOFIA Institut (DSI) under DLR contract 50 OK 0901 to the University of Stuttgart. Financial support for this work was provided by NASA through award 4$\_$0152 issued by USRA. T.H is funded by the National Research Foundation of Korea (NRF) grant funded by the Korea government (MSIT) through the Mid-career Research Program (2019R1A2C1087045).

\bibliographystyle{plain}
\bibliography{/Users/thiemhoang/Dropbox/Papers2/cites_paperApJ,/Users/thiemhoang/Dropbox/Papers2/cites_Books}

\end{document}